\documentstyle[12pt]{article}

\begin{document}
\textwidth 160 mm
\textheight 240 mm
\topmargin - 0.8 cm
\oddsidemargin -0.2cm
\evensidemargin -0.2cm
\headheight 0pt
\headsep 0pt
\topskip 9mm

\newcommand{\beq}{\begin{equation}}
\newcommand{\eeq}{\end{equation}}
\newcommand{\bq}{\begin{quotation}}
\newcommand{\eq}{\end{quotation}}
\newcommand{\BFACE}[1] {\mbox{\boldmath $#1$} }

\def\theequation{\arabic{section}.\arabic{equation}}

\title{ Why do we live in 3+1 dimensions? }

\author{
{\sc Holger Bech Nielsen} \\
{\sl The Niels Bohr Institute,} \\
{\sl Blegdamsvej 17, 2100 K\o benhavn \O, Denmark } \\
\and
{\sc Svend Erik Rugh\thanks{On leave from The
                  Niels Bohr Institute (1/1 - 1/7 1993). E-mail addresses:
                  ``HBECH@nbivax.nbi.dk'', ``RUGH@mitlns.mit.edu'' and
                  ``RUGH@nbivax.nbi.dk''}   } \\
{\sl Center for Theoretical Physics,} \\
{\sl Massachusetts Institute of Technology,} \\
{\sl 77 Massachusetts Avenue,} \\
{\sl Cambridge, MA 02139, U.S.A.} \\
}

\date{}
\maketitle

\setcounter{footnote}{0}

\begin{center}
{\small{\bf Abstract} }
\end{center}

{\small
Noticing that really the fermions of the Standard Model are best
thought of as Weyl - rather than Dirac -  particles
(relative to fundamental scales located at some presumably very
high energies)
it becomes interesting that
the experimental space-time dimension is singled out
by the Weyl equation:
It is observed that precisely in the experimentally true space-time
dimensionality 4=3+1 the number of
linearly independent matrices $n_{Weyl}^2$
dimensionized as the matrices in the Weyl equation equals the dimension
$d$. So just in this dimension
(in fact, also in a trivial case $d=1$)
do the sigma-matrices of the Weyl-equation form a basis.
It is also characteristic for this dimension that there is no
degeneracy of helicity states of the Weyl spinor for all nonzero
momenta.

We would like to interpret these features to signal a special
``form stability'' of the Weyl equation in the phenomenologically true
dimension of space-time.
In an attempt of making this stability
to occur in an as large as possible basin of
allowed modifications we discuss whether it is possible to
define what we could possibly mean by ``stability of Natural laws''.
}





\newpage

\setcounter{page}{2}

\section{Introduction}

Since ancient times \cite{BarrowTipler}
it has been a challenging question why we have
just three space dimensions. In the light of special relativity one
may formulate an even slightly extended question: Why do we have just
3+1 dimensions, one time and three space dimensions?

It is the purpose in this contribution to elaborate on
the idea \cite{Scottish} that the number of dimensions 3+1
is connected with properties of the Weyl equation, the latter
being especially ``stable'' in these dimensions (stability under
perturbations which do not keep Lorentz invariance, say, a priori).

In this connection
we shall however also
challenge the concept of ``stability of Natural laws''
(so much studied by one of us \cite{Scottish}, \cite{FroggattNielsen}).
This (second) part of our article is a preliminary effort where
we try to seek some ``principal boundaries''
for the set of ideas which we call the
``random dynamics project''.

Before we drift into a detailed discussion of these issues, i.e. our
own points of view, which thus aims at connecting the space-time dimensionality
with certain properties of the Weyl equation, we would like to
mention some other ideas
on how to arrive at the space-time
dimensionality 3+1. Ideas which  to some extent
are competing (cf. final
remarks, sec. 4.1.) with our own point of view taken here.

Let us first remark, that one of the purposes of this article is to
emphasize the scientific value of reading
information from phenomenology, i.e. from theories which have been
successfully confronted with experiments. Such inspiration from
phenomenology supplement more wild and speculative constructions
(theories of ``everything''?\footnote{We claim, however, that
in a certain elementary sense of the word ``everything''
it is not (theoretically) ever possible to arrive at a
theory of {\em everything} since an ``infinite regress'' argument
very fast sets in - which stops, only, if language itself (which parts
of it?) unambiguously could
select a unique construction of a ``T.O.E.''.
Not even then, however, does our search for a ``truth behind''
stop, since we are then, obviously, forced into a philosophy of
the (mysterious) power of our language. H.B.Nielsen and S.E.Rugh, ongoing
discussions. See also S.E. Rugh \cite{SvendetalNOGO}. }) which,
in case they do not a priori operate in 3+1 dimensions, clearly have
to offer some explanation for  why our space-time is 3+1 dimensional.

Phenomenologically inspired attempts of understanding
3+1 dimensions have indeed appeared, cf. e.g. P. Ehrenfest \cite{Ehrenfest},
G.J. Whitrow \cite{Whitrow} and the book by J.D. Barrow \&
F.J. Tipler \cite{BarrowTipler}.
The arguments presented in this context seems dominantly to be
based on observations involving
the Coulomb or Newton potential, connected to considerations
dealing with electromagnetic and gravitational interactions, respectively.
Assuming a Laplace equation for the (electromagnetic or gravitational)
field, it has a power related to
the dimension of the space. Thus, it can easily be calculated that
the Laplace equation in $d_s$ independent variables,
$ \bigtriangleup \Psi (x_1,...x_{d_s}) = 0$, will correspond to an
inverse $(d_s-1)$th power law of force (i.e. an inverse $(d_s-2)$th power
law of the potential) in a Euclidian space of $d_s$ dimensions.
Moreover, this power in the potential is related to
the stability of atoms or planetary systems. If the numerical
value of the power is larger than just the unity
corresponding to four dimensions -
the atom or planetary system gets unstable against falling into the singularity
at $r=0$ (or it goes  off to infinity, cf. P. Ehrenfest \cite{Ehrenfest}).
This instability concerns an unstable {\em state} in the terminology
of subsection 3.1 below. The argument being the main study of this
article turns out to be based on stability in a different sense:
Stability against variations in the {\em dynamics} (rather than the state).

Other - but more speculative - attempts have been made to explain
that there should be 3+1 dimensions: The following list is,
we admit, highly selective and random -
influenced by talks and papers which we accidently have come across.
G.W. Gibbons \cite{Gibbons} try to get 3+1 dimensions out of
a membrane-model in a higher dimensional setting;
K. Maeda \cite{Maeda} study how to get 3+1 dimensional universes as attractor
universes in a space of higher dimensional cosmologies and
a  group of physicists and astrophysicists in
Poland, M. Biesiada, M. Demianski, M. Heller, M. Szydlowski and J. Szczesny
\cite{Demianski}, pursue the viewpoint of dimensional
reduction - i.e. arriving at our 3+1 dimensional universe
from higher dimensional cosmologies.

In Copenhagen, pregeometric models have recently been investigated
by e.g. F. Antonsen \cite{Antonsen}. He arrived at the spatial
dimension $d_{rm \; eff} = 42/17 \approx 2.5 $ from a pregeometric
toy-model for ``quantum gravity'' based on random graphs, but according to
Antonsen this value is probably too low, since the model was only tested
with simplexes of dimension $0-4$. E. Alvarez, J. Cispedes and
E. Verdaguer \cite{Alvarez} also
arrives at dimensionalities around $d \approx 1.5 - 2.5$ in pregeometric
toy-models assigned with a (quantum) metric described by random matrices.

D. Hochberg and J.T. Wheeler \cite{HochbergWheeler}
contemplate whether the dimension may appear
from a variational principle. Also Jeff Greensite \cite{Greensite}
uses a variational principle (roughly making the signature
of the metric a dynamical variable) in order to get
especially the splitting 3+1 into
space and time.

Of course, any development sensitive to dimension and working in the
experimental dimension 3+1 may be considered an explanation
of this dimension, because an alternative dimension might
not be compatible with the same theory. For example, twistor theory
(cf. R. Penrose \cite{Penrose}) is such a theory, suggesting the dimension.
In fact, it does it in a way exceptionally close to the
``explanation'' presented in the present article, since both
are based on Weyl-spinors.

\vspace{1.0 cm}

\noindent
{\bf How superstrings arrive at 3+1 dimensions ?}  \\
Superstring theory \cite{strings}
has been much studied for a decade as a candidate for a fundamental theory.
Let us therefore finally discuss - more lengthy
- how the superstring, by construction living in 10 or 26
dimensions, may arrive at our 3+1 dimensional spacetime.

The space-time dimension $3+1$ (on our scales)
is in superstring theory singled out by several, to some extent, independent
considerations:

{\bf (1)} An argument for the dimensionality $d \geq 3+1$
(cf. R. Brandenberger \& C. Vafa \cite{Vafa}):
If one assumes some initial beginning - a big bang, say - with a lot
of superstrings wound around various dimensions of space
(think for simplicity of a torus) there should statistically
be strings of opposite orientations normally enclosing the same
dimensions, and they would ``as time passes'' compensate each other
and disappear provided they manage to hit each other and switch
their topology, so that such compensation is made possible.
As long as there are less than 3+1 (= 4) large (i.e. extended) dimensions
there is a very high chance that strings will hit and
shift topology  because it is very difficult
for strings in 3 or less space-dimensions
to avoid hitting each other.
It turns out that it is basically the number of
large dimensions, that
counts for whether ``hits'' take place easily or not.

As the dimension of the large directions reaches 3+1 the process of
unfolding slows drastically down and one may imagine this being
the reason for there just being 4 dimensions.
We remark, that this argument is truly very ``stringy''
since it makes the dimension of space-time become just twice
that  of the time track of the string (i.e. $ 2+2 = 4 = 3+1$).

{\bf (2)} A compactification argument (``dimensional reduction''):
A Calabi-Yau space has exactly $6$ dimensions, and so the
$10 = 9+1$ dimensions needed for the superstring to exist (because of
the requirement of cancellation of anomalies,
or, say, to get the needed zero-point mass squared - contribution as needed)
enforces that using $6$ of the $10$ dimensions
in the compactification leaves just $4 = 3+1$
for the ``large'' dimensions
But, why should
we select a Calabi-Yau space as the compactifying space?
It is  motivated by assuming that $N=1$ supersymmetry
should survive far below the compactification scale of energy, i.e.
by the requirement of not breaking this supersymmetry.

While the argument using Calabi-Yau space and the 10 dimensionality
of the superstring can hardly stand alone without
other support for the string theory, several of the other
arguments are sufficiently primitive to be trusted by itself, since
based on phenomenologically supported ideas. Note, for instance, that
argument {\bf (1)} for string theory above was also to
some extent phenomenological.

Let us now, for a moment, believe in the compactification mechanism, reducing
the dimensionality from 10 to 4=3+1. Then, it becomes a central issue
of clarification at which scales this compactification takes place.
It is challenging for the superstring that it has
been suggested by Antoniadis \cite{Antoniadis} that already at a few TeV one
should expect to see the additional dimensions. If true, this
has
- most likely - severe cosmological implications.
One of us \cite{Antoniadis} would like to make the arguments of Antoniadis
completely rigorous, so that strings can not escape this prediction
(modulo very general assumptions about string-constructions).

It is interesting, though, that recent attempts
tend to formulate superstring theories from the
beginning in four spacetime dimensions, rather than existing in 10 or
26 dimensions, of which all but the four extended dimensions of our
spacetime somehow become compactified. (Cf, e.g., the summary talk by
Steven Weinberg \cite{Weinberg} and references therein).

But in that case the superstring
has evidently a somewhat less impressive
capacity of power as regards
an explanation of why we live in 3+1 dimensions.

\subsection{Getting inspiration from phenomenology}

Instead of relating the dimension number ``3+1'' to highly
speculative constructions, which could be plain wrong, such as
strings, we would
rather try to identify structures in the well
known and established laws of Nature which point towards that the
number ``3+1'' is distinguished.

You may consider this project a sub-project of a program consisting
in {\em ``near reading''}
\footnote{The desperation in this project is that all
this information can be immensely condensed down (to a page or so) so
we have actually only very little  structure to ``read''
(and to be inspired from): The truly observed sector of the
Standard Model has only of the order of 19 real parameters,
some of which are hardly determinable in practice, even using
baryon number generation physics: 12 quark and lepton
masses of which the three neutrino masses (in the simplest
Standard Model) are understood to be just zero,
three finestructure constants,
four parameters in the Kobayshi-Maskawa-Cabibbo matrix and
the $\theta$-angles.

In this program of ``near reading'' information from the Standard Model
we have also tried to read off  remarkable
features  of the Standard Model gauge group $S(U_2 \times U_3)$
compared to other groups with up to 12 generators (12 gauge
bosons), say, and we have
found that the standard model group is remarkable by its
{\em ``skewness''} (remarkably few automorphisms), cf.
H.B. Nielsen \& N. Brene \cite{skewness}. }
of the empirically based structure we know
today, i.e. the formal structure and the parameters of
the Standard Model (describing the electroweak and strong
``spin 1'' gauge-interactions) and Einsteins theory for the gravitational
``spin 2'' interactions.

Among the information we do not understand today (and which we necessarily
have to be ``inspired'' from) the following issues were especially
stressed by Steven Weinberg \cite{Weinberg}
at the recent XXVI Conference on High Energy Physics in Austin
(August 1992): \\

\noindent
$\bullet$ Why do the parameters of the Standard Model
take the observed values ? \\
$\bullet$ Why are there three generations of quarks and leptons ? \\
$\bullet$ Why is the Standard Model group $S(U_2 \times U_3) \simeq
U(1) \otimes SU(2) \otimes SU(3)$ ? \\

\noindent
And here we would like to add: \\

\noindent
$\bullet$ Why do we live in 3+1 dimensions ? \\

\noindent
In fact, we have found a special property in the
Weyl equation which pervades so strongly the
Standard Model of the electroweak and strong forces and the
known material constituents.\footnote{Note, that even relative
to the $W$ mass scales most of the quarks and leptons are light
particles, and certainly they are all ``massless'' relative to
the Planck scale. Since also right and left components couple
differently to the weak gauge bosons, a description in terms of
Weyl particles is strongly suggested, cf. also introductory
discussion in H.B. Nielsen \& S.E. Rugh \cite{HolgerSvend}. }
This special property (which we encode by
the concept of ``stability'' and
which we shall discuss in this contribution)
is a property which distinguishes 4=3+1 and 1=0+1
as standing out relative to all other space-time dimensionalities.
The number of linearly independent matrices of the type
that appear in the Weyl equation is exactly equal
to the dimensionality
$d$ of space-time if $d=4$ (or, in fact, if $d=1$).

Are the elementary matter constituents, we know today (i.e. the fermions)
not Dirac particles rather than Weyl particles?
Well, the Weyl spinors - introduced by Hermann Weyl -
were originally rejected as candidates for
our constituents of matter
because they were incompatible with parity conservation.
However, seeking fundamental physics (at the Planck scale, say) we ignore
the small masses of a few hundred $GeV/c^2$ of the quarks and the
weak gauge bosons, rendering the connection between right and left handed
components completely negligible. Thus we consider all the material
constituents (quarks and leptons) we know today
as Weyl particles rather than Dirac particles!

So to read some special feature of dimension 3+1 connecting to the
Weyl equation means to read it off from
all known material fields of which we are build up!
(cf., also, the more detailed introductory discussion
in H.B. Nielsen \& S.E. Rugh (1992) \cite{HolgerSvend}).

The outline for our contribution is as follows:
In the following section (section 2) we \
remind the reader about the Dirac
and Weyl equations in an arbitrary number of dimensions.
Especially, we calculate the number of components
of the Weyl spinor and the crucial observation of
the equality of dimension and the number of elements in a basis for
the matrices in the Weyl equation is made. In section 3 we then seek
to unravel the message to be learned from this observation: It is connected
with ``a stability'' under perturbing the fundamental ``dynamics''.
We also discuss the general limitations for postulating such a stability.
In section 4 we resume and put forward the concluding remarks and dreams,
and it is discussed if the observation could just be an accident?

\section{The Dirac and Weyl equations in $d$ dimensions}

\setcounter{equation}{0}

In order to note the speciality of the
4=3+1 dimensional Weyl equation we shall see that in the general
$d$ dimensional case the number of components $n_{Weyl}$ of the
spinor $\psi$ of what is reasonably called the Weyl
equation
\begin{equation}\label{WE}
 \sigma^{\mu} D_{\mu} \psi = 0
\end{equation}
is
\begin{equation}
n_{Weyl}=\left\{\begin{array}{cc} 2^{d/2-1}  &
\mbox{for $d$ even,}\\
2^{(d-1)/2}&  \mbox{for $d$ odd.}\end{array}
\right.
\end{equation}
In fact, the Weyl equation is at first glance only defined for even
dimensions and, thus, we have to agree upon, what we will call a Weyl
equation in odd dimensions. We { outline shortly} how
to build up a Weyl equation in even and in odd dimensions.

\subsection{Construction of the Weyl-equation (in even and odd dimensions)
from the Dirac-equation (in even dimension)}

The Weyl-equation in arbitrary dimensions is most easily constructed from
the Dirac equation, which we shall in the next section construct
in even dimensions via an inductive construction. It  has only half
as many components and is obtained by ``$\gamma^{d+1}$-projection''
(think here of the $\gamma^5$-projection in the 4 dimensional
case) meaning that  we first define
\begin{equation}
\gamma^{d+1}=\gamma^0\gamma^1\gamma^2 \cdot \cdot \cdot \gamma^{d-1}\gamma^d
\end{equation}
modulo an optional extra phase factor $i$ so as to be hermitean,
then choose $\gamma$-matrix
representation so that this $\gamma^{d+1}$ becomes diagonal
(Weyl-representation), and then thirdly
write only that part of the Dirac equation
\begin{equation}
(i\gamma^{\mu}D_{\mu}-m)\psi =0,
\end{equation}
 which concerns those components of the Dirac field $\psi$ which
correspond to one of the eigenvalues of $\gamma^5$, say $+1$.
We now drop the mass-term. Since the only remaining term in the Dirac-equation
is thereby the ``kinetic'' one $i\gamma^{\mu}D_{\mu} \psi =0$ and the
gamma-matrices only have matrix elements connecting
$(\gamma^{d+1}=1)$-components
to $(\gamma^{d+1}=-1)$-components the Weyl equation can be written
using only matrices with half as many components as the corresponding Dirac
equation from which it is obtained. Really we find the Weyl matrices
as off-diagonal blocks in the Dirac matrices in the Weyl representation.
That is to say the Weyl-gamma-matrices which we call $\sigma^{\mu}$
(or $\sigma^{\mu}_{-}$ and $ \sigma^{\mu}_{+}${\bf )} if we want to distinguish
if we projected on the $\gamma^5$ equal to minus or plus 1 components)
are given by
\begin{equation}
\sigma^{\mu}_{-}=\gamma^{\mu}\frac{1 - \gamma^5}{2} \; \; \; , \; \; \;
\sigma^{\mu}_{+}=\gamma^{\mu}\frac{1 + \gamma^5}{2}
\end{equation}
with successive removal of the unnecessary entries in the matrices, or
better by
\begin{equation}
\gamma^{\mu}=\left( \begin{array}{cc}
0 & \sigma^{\mu}_{+} \\
\sigma^{\mu}_{-} & 0
\end{array}\right) .
\end{equation}

\subsection{Inductive construction of the $\gamma$ matrices (by addition
of two dimensions at a time)}

If we are in $d_t$ time dimensions and $d_s$ space dimensions we denote
by relativistic invariance the invariance of the equations
(to be constructed right now) under
the (generalized Lorentz) group $O(d_t,d_s) \equiv O(d_t, d - d_t)$.

Generalizing the construction by Dirac, we seek a construction in
$d = d_t + d_s$ dimensions, which involves a set of $\gamma$ matrices,
$\gamma^1, \gamma^2, .... ,\gamma^d $
shaped to obey a specific anti-commutator algebra
\begin{equation} \label{anti-commutator}
\{ \gamma^{\mu},\gamma^{\nu} \}=
2 g^{\mu\nu}_{(d)}
\end{equation}
(where $g^{\mu\nu}_{(d)}$ denotes
the metric tensor, taken just to be a series of $1$'s and $-1$'s
along the diagonal).

If equation (\ref{anti-commutator}) is satisfied for the $\gamma$ matrices,
the Dirac equation imply a corresponding Klein-Gordon equation
\begin{equation}
(i\gamma^{\mu}D_{\mu} + m) (i\gamma^{\mu}D_{\mu} - m) \psi =
(-g^{\mu\nu}D_{\mu}D_{\nu}-m^2
- [\gamma^{\mu},\gamma^{\nu}]eF_{\mu\nu})\psi  = 0
\end{equation}
(with magnetic moment term in it, and
where we used $ [D_{\mu},D_{\nu}]=2eF_{\mu\nu}$)
and that is what is wanted to have the usual relativistic
dispersion relation
$$ p^2-m^2=0 \; \; \; .$$
It is not difficult to show the existence of such $\gamma$ matrices
by induction in steps of two in the dimensions, starting, e.g.,
from two dimensions, to get the even ones.

\subsubsection*{Start of the induction in two dimensions}

In two space-time dimensions one may use two of the three Pauli matrices as
gamma-matrices, e.g.
\begin{equation}
\gamma^0=\sigma_x \mbox{    and    } \gamma^1=i\sigma_y
\end{equation}
in the Minkowski-space case and simply
\begin{equation}
\gamma^0=\sigma_x \mbox{    and    } \gamma^1=\sigma_y
\end{equation}
in the euklidian(ized)
two dimensional space. These are the Dirac-matrices
and they have $n_{Dirac}=2$ components.

\subsubsection*{Step from $d-2$ to $d$ dimensions.}

Having the Dirac equation in $d-2$ dimensions (with $d$ even)
we can in an inductive construction make the Dirac-matrices for
two dimensions higher by the following construction:

Suppose that we already have constructed the gamma-matrices for
$d-2$ dimensions and denoted them with a tilde in order to distinguish
them from the gamma-matrices for $d$ dimensions which
we construct successively. They obey the anti-commutation algebra
\begin{equation}
\{ \tilde{\gamma}^{\mu},\tilde{\gamma}^{\nu} \}=2g^{\mu\nu}_{(d-2)}
\end{equation}
with
\begin{equation}
\mu,\nu=1,2,3,...,d-3,d-2.
\end{equation}
Then we define the $d$-dimensional gamma-matrices for the first $d-2$
values of $\mu$ by:
\begin{equation}
\gamma^{\mu} = \left( \begin{array}{cc}
0 & \tilde{\gamma}^{\mu}\\
\tilde{\gamma}^{\mu} & 0  \end{array} \right)
\label{ind1}
\end{equation}
and the two new gamma-matrices by:
\begin{eqnarray}
\nonumber \gamma^{d-1}=\mbox{\it sign } \cdot \left(
\begin{array}{cc} 1 & 0 \\ 0 & -1
\end{array}\right) ;\\
\label{ind2}\\
\nonumber \gamma^{d}=\mbox{\it sign } \cdot \left(
\begin{array}{cc} 0 & -i1 \\ i1 & 0
\end{array}\right).
\end{eqnarray}
The signs {\it sign} (which, without loss of generality, need only to
take values $1$ or $i$) are to be chosen so as to
arrange the wanted signature
for the anti-commutator algebra\footnote{In a systematic exposition of
this construction one would go into Clifford algebras and all that.}
\begin{equation}
\{ \gamma^{\mu},\gamma^{\nu} \}=2g^{\mu\nu}_{(d)}
\end{equation}
which is easily checked with the presented ansatz, cf. (\ref{ind1})
and (\ref{ind2}).

\subsection*{Number of components of the Weyl-and Dirac spinors}

Thus we see that the number of components for a Dirac equation
for even dimension $d$ of space-time is\footnote{This number of
components is only a minimal number
in the sense that the Weyl spinors could have additional degrees of
freedom (as they indeed have, for instance ``color''). When we have - until
now - ignored such additional degrees of freedom, this is, mathematically
speaking, that we have build up the irreducible representations of the
algebra (\ref{anti-commutator}). The question about additional
degrees of freedom of the Weyl spinor turns out to be a major
``killing attempt'' of the entire approach pursued here,
cf. the section ``An almost true theorem ...'' (subsection 3.6). }
\begin{equation}
n_{Dirac}=2^{d/2}       \mbox{    for $d$ even}
\end{equation}
and thus since the Weyl equation has just half as many components it
has:
\begin{equation}
n=n_{Weyl}=2^{d/2-1}   \mbox{    for $d$ even}
\end{equation}
Note, that in $d=4$ dimensions the Weyl spinor gets two components
(the neutrino), while in 10 dimensions, say, the Weyl spinor gets
16 components (a complicated neutrino, indeed) so we see that the
number of components grows very fast, being an exponential
function of the dimension $d$. I.e. whereas a ``spin 1/2'' particle
has 2 spin-states (if it is a massive Dirac spinor) as we are used to
in four space-time dimensions, a ``spin 1/2'' particle in 10
dimensions has 16 spin-states (but 32 Dirac-components).

\newpage

\subsection{The odd dimensional cases}

In odd dimensions one may simply include the $\gamma^{(d-1)+1}$-matrix as the
matrix number $d$ and use the $d-1$ even dimensional Dirac equation, which
we have already constructed. That is to say, we take the Dirac equation
one unit of dimension lower than the odd-dimensional equation we want to
construct. Then we have all the $\gamma$-matrices already except for the
last one - the $d$th one. For the latter we then use the construction of
taking the product of all the first $d-1$ matrices. Since it is well known that
such a ``$\gamma^5$''-type product will anti-commute with all the other
gamma-matrices if constructed for even dimension, and since it can by an
optional factor $i$ have the sign of its square as wanted, it obviously
will satisfy all the needed properties for joining the set of
gamma-matrices to get a set for the odd dimension $d$.

The equation, with gamma-matrices,
obtained in this way for the odd dimension
$d$ is in two ways best considered
the odd-dimensional{\em  Weyl} equation:

{\bf 1.} It violates parity\footnote{In odd
space-time and thus even space dimension the point-reflection
is just a rotation and when we talk about parity we mean a reflection
in a mirror (true mirror symmetry).},
just like the even-dimensional Weyl equation,

{\bf 2.} It cannot be further reduced by ``$\gamma^5$''-projection
(really $\gamma^{d+1}$),

since its ``gamma-five'' would become the unit matrix.  \\
However, this ``Weyl equation'' in odd dimensions deviate from the Weyl
equation properties by
allowing a mass term, since there is only a trivial
$\gamma^{d+1}=1$ that can be constructed.
There is no ``chirality'' which protects against
generation of mass $m \neq 0$ for the Weyl spinor field $\psi$.

One might then define an equation with the double number of components
and consider {\em that} the odd dimensional Dirac equation. By composing a
couple of odd dimensional Weyl equations, being mirror images of each other
one easily obtains a {\em ``Dirac equation''} in odd dimensions
which is both symmetric under parity transformations
and can be reduced back to its Weyl-components.

With the suggested notation we easily find the number of components
for the odd dimensional case:
\begin{equation}
n_{Dirac} = 2 \cdot 2^{(d-1)/2}=2^{\frac{d+1}{2}} \mbox{    for $d$ odd}
\end{equation}
for the Dirac equation (the doubled one) and for the
Weyl equation
\begin{equation}
n_{Weyl} = 2^{(d-1)/2}=2^{\frac{d-1}{2}} \mbox{    for $d$ odd}.
\end{equation}

\subsection{An observation}

With $n_{Weyl}$ components the number of linearly independent
$\sigma^{\mu}$ type matrices that can be formed is $n_{Weyl}^2$.
Out of these $n_{Weyl}^2$ possibilities the Weyl equation makes use
of $d$. Our crucial observation is that in the phenomenologically true case
$d=4=3+1$ it happens that all the $n_{Weyl}^2$ possible matrices
are {\em just used once each} since in fact
\begin{equation} \label{observation}
n_{Weyl}^2 = d.
\end{equation}

This equation written as an equation for the dimension $d$ of
space-time becomes
\begin{equation}\label{insertobservation}
d=\left\{\begin{array}{cccc}( 2^{d/2-1})^2 & = & 2^{d-2} &
\mbox{for $d$ even,}\\
(2^{(d-1)/2})^2 & = & 2^{d-1} & \mbox{for $d$ odd}.\end{array}
\right.
\end{equation}
As can be easily seen from the figure the only acceptable solutions to
this equation are
\begin{equation}
d=\left\{ \begin{array}{c}1\\4 \end{array} \right.
\end{equation}
because the solutions appearing at first also $d=2$
(if $2$ had been odd)
and an irrational solution around $d\approx 0.3$
(had it been even) are not acceptable.

\begin{figure}
\vspace{12.0 cm}
{\em Fig.1. This figure shows the graphical way of solving equation
(\ref{insertobservation}) by presenting as functions of the
dimension $d$ a semilogarithmic plot of : \\
1) the identity function $d$ (the curved curve),
2) right hand sides of (\ref{insertobservation}) interpolated to
real numbers separately for even and odd $d$ (the two straight lines).
Note, that there is also coincidence for $d=1$ which is however a completely
empty and trivial theory.
The one for $d \approx 0.33$ is of course not a true solution.
}
\end{figure}

\newpage

\section{The equation $n_{Weyl}^2 = d$ as a message about
robustness and stability of the Weyl equation in 3+1
dimensions}

\setcounter{equation}{0}

What could possibly be the reason for this coincidence of numbers
$n_{Weyl}^2$ and dimension $d$ ? (If it is not an accident, cf. the
final remarks).

It means that the sigma-matrices $\sigma^{\mu}$ make up a basis
for all the possible matrices that could possibly
multiply the Weyl spinor in $d$ dimensions, if $d=4$
(or in $d=1$, which is a very trivial case
indeed \footnote{Considering this one dimension
a time dimension, there is a zero dimensional space or, in other
words, only one point.
The Weyl equation has in this case the following
trivial form, $D_0 \psi = (\partial_t - i e A_0) \psi = 0$, where $A_0$ can
be gauged away, rendering the single-component Weyl spinor
completely constant. We thus have a Hilbert space
with only one dimension, and thus
there is only one quantum state for the Weyl spinor, leaving no
room for developments in that ``universe''.}). In particular,
this means that
$$ \sigma^{\mu} D_{\mu} $$
may be considered as the most general linear operator that can act
on $\psi$, provided we could consider the $D_{\mu}$'s completely
general expansion coefficients.
Now, in fact, $D_{\mu}$ is the covariant
derivative $D_{\mu} = \partial_{\mu} - i e A_{\mu}$. So is this
really the most general operator ?
Yes, it is indeed the most general form of the first two terms in a
Taylor expansion of any reasonable
well behaved differential operator
\begin{equation}   \label{TaylorOmega}
\Omega (x,p)
\approx
\Omega (x, p = 0) +
\frac{\partial \Omega (x, p = 0)}{\partial p_{\rho}} p_{\rho}
+ \frac {1}{2}
\frac{\partial^2 \Omega (x, p = 0)}
{\partial p_{\rho}   \partial p_{\sigma} } p_{\rho} p_{\sigma} + ....
\end{equation}
The most general equation, which we suppose to be
homogenous and
linear in the Weyl-spinor field $\psi$ (see discussion later), is thus
\begin{equation} \label{Weylkandidat}
 \Omega (x,p) \psi (x) = 0  \; \; , \; \;
p = p_{\mu} = i \partial_{\mu}
\end{equation}
where now $\Omega$ denotes an $2 \times 2$ matrix operator, where we
have considered only operators $\Omega$ which map Weyl-spinor waves
$\psi (x) = \psi (\vec{\BFACE{x}},t)$ to Weyl-spinor waves (as in the
Weyl equation).

Being a $2 \times 2$ matrix operator, $\Omega$ can be expanded in terms
of the basis matrices $\sigma^{\mu}$ (in the case where the
equation (\ref{observation}), i.e. $n_{Weyl}^2 = d$, is fulfilled
we can use the
set of the matrices $\sigma^{\mu}$ as a basis).
Let ${\Omega}_{\mu}$ denote the basis coefficients, i.e. define
\begin{equation}
\Omega (x, p)  = \Omega_{\mu} (x,p)  \sigma^{\mu}
\end{equation}
We Taylor expand:
\begin{eqnarray}   \label{TaylorD}
{\Omega}_{\mu} (x,p) & =  & {\Omega}_{\mu} (x, p = 0) +
\frac{\partial {\Omega}_{\mu} (x, p = 0)}{\partial p_{\rho} } p_{\rho}   +
...   \\
 & = & V_{\mu}^{\lambda} e A_{\lambda} (x) + V_{\mu}^{\rho} (x) p_{\rho}
+ ...
\end{eqnarray}
In the last equation (which is merely notation) we have defined
\begin{equation}
V_{\mu}^{\lambda} (x) =
\frac{\partial}{\partial p_{\lambda} }
\Omega_{\mu} (x, p=0) \; \; \; ;
\; \; \;
e A_{\lambda} (x) = V^{-1 \; \mu}_{\; \; \lambda}\; \Omega_{\mu} (x,p=0).
\end{equation}

By identifying $A_{\lambda}(x)$ with an electromagnetic field and
$V_{\mu}^{\lambda} (x)$ with a vierbein (in a gravitational theory)
we have thus made the interpretation of  the lowest order terms in the
Taylor expansion of the general $2 \times 2$ matrix operator equation
(\ref{Weylkandidat}) as being nothing but
the usual Weyl equation for our Weyl-spinor $\psi$ (coupled
to an electromagnetic field $A_{\lambda}$ and to
a gravitational field in the usual manner by the vierbein field
$V_{\mu}^{\lambda}$), see also discussion in section 3.5.

We can summarize that Nature has organized
just such a dimension $d$ of space-time
as to make  the Weyl equation operator (with couplings to
some external electromagnetic and gravitational fields
unavoidably appearing in the same go)
the most general operator
in the Taylor expanded limit of small momenta
$p=p_{\mu}$ (small derivatives
$\partial_x  \psi \ll \psi $).

Why do we need all these Taylor expansions? Well,
that we arrive at the Weyl equation via Taylor expansions (keeping only
the lowest order terms) is a very natural thing, bearing in mind that
we are living in the ``infrared limit'' compared to
some presumed fundamental scale (e.g., the Planck scale, say)
in the sense that
we have - even with our best accelerators today -
access to very small 4-momenta $p$ only.

In fact, to use such a Taylor expansion (at experimentally accessible energies)
is completely  analogous to
Taylor expanding away higher order curvature terms in the gravitational
action thereby arriving at the Einstein-Hilbert action in the ``low
energy limit''. And just like higher order curvature terms will
blow up in the ultraviolet in the action for the gravitational field
we could
expect that the Weyl equation will be modified in the
ultraviolet (for higher momenta) and blow up terms
quadratic in the momenta like
$\bar{\psi} \sigma^{\mu} \partial_{\mu} \partial_{\nu} \psi \xi^{\nu}$
(here $\xi^{\nu}$ denotes some vector field,
which do not necessarily, in fact better
not, have to be the $A^{\nu}$ field) etc. etc

The above means, especially, { that we can make a reinterpretation
of any little additive change}
\begin{equation}   \label{Weylperturbation}
\sigma^{\mu} D_{\mu} \psi = 0  \; \; \; \rightarrow \; \;  \;
(\sigma^{\mu} D_{\mu} + \delta W) \psi = 0
\end{equation}
in the operator $\sigma^{\mu}D_{\mu}$ acting on $\psi$
as a change in the $D_{\mu}$'s which again is just a
shift in the ``electromagnetic field'' $A_{\mu}$ and/or the
``gravitational'' vierbein $V_{\mu}^{\lambda}$.

Here $\delta W = \delta W (x,p) $ could be any $2 \times 2$
matrix operator, which have a well behaved Taylor expansion in such
a way that higher order terms than those which are linear in
4-momenta $p$ may be neglected when $p$ is small.

Notice that the type of modification terms $\delta W (x,p) $
we here consider are not at all Lorentz invariant a priori.
That the Weyl equation keeps its form and thus Lorentz invariance
is only achieved at the cost of allowing the vierbein $V_{\mu}^{\lambda}$
(and $A_{\mu}$) to be modified, but it is still remarkable.

We would thus like to interpret
the dimension coincidence
as a signal of a
certain kind of ``form stability'', which
we are going to discuss further in a moment. \\

\noindent
{\bf In 4 dimensions there is only one helicity state (no
degeneracy) of the Weyl spinor:} \\
An alternative way, to observe the ``stability'' of the Weyl spinor
in exactly 4 dimensions is to notice that a Weyl {\em particle}
(antiparticle not included !) for a given momentum has just
the following number of Weyl particle polarization states
\begin{equation}
\frac{n_{Weyl}}{2} = \left\{\begin{array}{cc} 2^{d/2-2}  &
\mbox{for $d$ even,}\\
2^{(d-3)/2}  & \mbox{for $d$ odd}.\end{array}
\right.
\end{equation}
which means that it is just for $d = 3$ or $d = 4$ that there is just
no degeneracy of particle states and just one state per momentum!
Several states for one momentum signals an instability under
modifications $\delta W (x,p) $, because in general such a
modification will cause level repulsion (of the eigenvalues),
as is well known from
perturbations of levels in quantum mechanics.
\footnote{E.g. for a 6-dimensional Weyl equation,
which according to the observation
above, has degeneracies of states, this degeneracy would disappear
under (almost) any modification, and hence the modified (perturbed)
equation could not be interpreted as a 6-dimensional Weyl equation. }
We can (at least) say
that $d=4$ is the highest dimension without degeneracy!


\subsection{Is it possible to define ``structural stability'' for
Natural Laws? }

Let us for a moment set aside the Weyl equation (we return to that)
and digress into a discussion of how  we  possibly can
assign a meaning to the notion of ``stability of Natural laws''
(and, in particular,
the stipulated stability of the Weyl equation).

It is remarkable that there are
over fifty
generally accepted definitions
of stability (which we shall not go into here,
cf, e.g., Szebehely (1984) or E. Atlee Jackson, p.41
\cite{AtleeJackson} ). These definitions
deal, however,
with the stability of a certain state, i.e. a specific configuration of the
degrees of freedom of a fixed dynamical system, a fixed Hamiltonian, say.

The Weyl {\em equation} is a law of Nature (or a
dynamics), not a {\em ``state''}.
So let us distinguish the
concept of stability of a ``state''
under perturbations in the space of possible states (with the Natural
laws fixed) from that of  stability of the laws (the dynamics)
under perturbations in the space of
possible laws of Nature (possible Hamiltonians, say):  \\

\newpage

\subsubsection*{I. Stability of states (keeping the dynamics, i.e.
the Natural laws, fixed)}

Taking into account the ubiquitous
bath of small perturbations
and ``noise'' which surrounds
and ``attacks''
every object (in some  specific ``state''), it necessarily has
to possess some degree of stability (along the directions of the most likely
perturbations which ``bath'' the object)
in order to exist in that ``state'' for a longer interval of time.

This is the type of stability that distinguishes
a needle standing vertically on its tip as unstable
versus a stable ball,
say, lying in a (little) depression. To make
the needle balance for a long interval of time
requires enormously accurate fine tuning, although it is in principle
possible.

\subsubsection*{II. Stability of a dynamics (a set of equations)}

Rather than dealing with the stability of a ``state''
we would
like to  focus on the ``stability of a dynamics''.
In this case, the questions  which we will address
are whether certain features of the dynamical development are
stable
under modifications of the Hamiltonian or action, say.

\subsubsection*{II(a) ``Structural stability'' of a dynamical system}

In particular, the concept of ``structural stability''
(first put forward by A. Andronov and L.S. Pontryagin more than fifty years
ago) of a dynamical system is defined in the following way:
\begin{quotation}
\noindent
A given physical model, based on the system
$\dot{\BFACE{\xi}} = \BFACE{F(\xi,c)} $
is "structurally stable" if a slight change in
$\BFACE{F}$,
$$ \BFACE{F} \; \rightarrow \;
\BFACE{F} + \delta \BFACE{F} \; \; \; \;, \; \; \; \;
|| \delta \BFACE{F(\xi)} || < \epsilon \; \; \; \;  \forall \BFACE{\xi} \in
\BFACE{R}^n $$
does not result in {\em "essential"} changes in the solutions of this
system, whereby is meant
that the phase portraits of $\dot{\BFACE{\xi}} = \BFACE{F(\xi)} $ and
$\dot{\BFACE{\xi}} = \BFACE{F(\xi)} + \delta \BFACE{F} $ are
{\em topologically} orbitally equivalent.
In other words, there exist homeomorphisms that transform the phase
space trajectories of the vector field
$\BFACE{F}$ to those of the perturbed one $\BFACE{F} + \delta \BFACE{F} $
(Cf., e.g., E. Atlee Jackson \cite{AtleeJackson} Vol.I, p. 102,
p.380. See, also, e.g.
A.A. Coley and R.K. Tavakol \cite{ColeyTavakol})
\end{quotation}

\subsubsection*{II(b) ``Stability of Natural laws''
(in the ``random dynamics'' spirit)}

Removing the focus from ``orbital topological equivalence''
we would attempt to consider as
the essential feature (the stability of which is to be defined)

\newpage

\noindent
some effective laws or regularities {\em appearing in some
limit} \footnote{Dominantly, we shall have
in mind this limit to be
an {\em infrared limit}, i.e. physics at our ``human scales'' where all
the particle and interaction constituents have small momenta and
energies compared to some ``presumed'' fundamental scale of the
``world machinery'' (this world machinery corresponds to the
system  $\dot{\BFACE{\xi}} = \BFACE{F(\xi)} $ of {\bf II.(a)} above).

But also other limits could be thought of: For instance the
isospin symmetry of strong interactions appears in the
``limit'' of both the up and the down quarks happening
to be very light compared to the scale given by
$\Lambda_{\overline{MS}}$.
We have, also, a proposal for understanding the linearity
and to some extend hermiticity of the Schr\"{o}dinger equation
as features appearing in the limit of  waiting very long,
i.e. very late times.
See, e.g., C.D. Froggatt \& H.B. Nielsen
\cite{FroggattNielsen}  }.  We may say
that we have stability in the spirit of ``random dynamics''
\cite{FroggattNielsen} if such effective laws or regularities are
unchanged under perturbations (``small'' modifications)
of the (fundamental) dynamics of the
system.

By analogy:
Just like a ``state'' of a given Hamiltonian, say,
have to possess a certain degree
of stability in order to exist for a longer span of time (otherwise
the perturbations have to be finetuned enormously), we speculate:
Could it be that the ``Natural laws'' also have to possess some degree of
stability ?
And how may one pursue this idea ?

If it {\em is} so that the Natural laws
\footnote{Down to scales of
$\sim 10^{-15}$ meter or so, the Natural laws
apparently do not change significantly
at different space-time points, since we seemingly have good
conservation of energy and momentum. It may be a bit hard
even to imagine how there could be a breaking of translation symmetry
at even smaller scales  - without causing problems
by being observable. But we may, either, e.g. imagine the coupling
to long waves to be only tiny \cite{FroggattNielsen} or some
sort of quantum fluid \cite{Lehtoetal}.
}
depend substantially on finetuned
implementations of structure at the Planck scale, say, then
there is very little ``room'' for
``mis-implementations'' and ``small errors'',
as regards the structure (the Natural laws) at this fundamental scale.

The belief that this is not the case, i.e.
the belief that the Natural laws indeed
will show up to exhibit some degree of stability in order to ``exist''
- that is the ``random dynamics spirit''! (and the viewpoint
have been investigated and implemented in the context of many different
``toy-models''.
Cf., e.g., C.D. Froggatt \& H.B. Nielsen
\cite{FroggattNielsen} and references therein.

However, the ``random dynamics'' point of view have inherently some
small difficulties and obstacles which we shall now seek to discuss.
(Section 3.2-3.5).

Note, also, that structural stability in the strict sense of {\bf II(a)}
demands complete equivalence
between phase portraits { whereas} ``the random dynamics''
spirit stability only  aspire to arrive at the same ``form''
in some ``limit'', where the same ``form'' means
that you can interpret it as following the same law(s).
Even though {\bf II(b)} is weaker requirement of stability
than {\bf II(a)} in the sense of
only caring for a limit, it is of course much stronger
if we take it, {\bf II(b)}, to mean stability of the Natural
laws under ``completely general'' perturbations (whatever that
should mean ?, cf. discussions below).

\subsection{How we have weakened the concept of stability in
order for the Weyl equation to be stable?}

Note especially, that in order to claim the stability of the Weyl
equation we {\em restricted } the class of perturbations
$\delta W$ (modifications) of that equation considerably:
We outline below in which sense this basin of perturbations was
restricted and comment for each item
how we imagine one could - to some extent - relax
the restrictions made.  \\

\noindent
{\bf 1. The equation is restricted to remain linear and
homogeneous in the Weyl spinor field $\psi$} \\
One may argue that additional
terms like $\psi^2$,$\psi^3$,...would be
suppressed if the $\psi$ field itself
is considered small. However there is a problem how to suppress
the zeroth order
term - i.e. the term not depending on $\psi$ -
in the limit  of weak $\psi$, unless one somehow argues
that { attention}
can be restricted to the
homogeneous part of the equation, the inhomogeneous solution
being just ``background''.  \\

\noindent
{\bf 2.} {\bf The number of degrees of freedom (of the Weyl spinor)
remained unchanged}  \\
We modified the Weyl equation for the
Weyl spinor $\psi$ (with two degrees of freedom) with
perturbations $\delta W$ restricted to be of the
operator-form of $2 \times 2$ matrices.

``Elementary constituents'' (elementary particles)
often turns out to be ``composite'' when they are
looked upon at higher energy scales (cf. the
 ``quantum ladder'').
I.e. they  { semi-always} turn up to have additional degrees
of freedom (``frozen in'' in the infrared limit)
which gradually ``wake up'' towards the ultraviolet.
That is, the effective number of degrees of freedom
of some object depends on the energy by which we ``look'' at it.

In particular, one could imagine a ``random dynamics'' project
where the  number of components of what
becomes the Weyl-spinor $\psi$ field itself (in the infrared)
did dependent on the energy-scales.

Indeed, the toy-model for a ``world machinery'' described in
detail in
section 3 of H.B. Nielsen \& S.E. Rugh \cite{HolgerSvend}
is to be considered as one
possible attempt of relaxation on exactly this point. It is
in fact argued that - genericly - the
Fermi surface would consist
mainly of points of the dispersion relation with at most
two degenerate states, because the points with more than
two cannot separate filled and unfilled states for topological
reasons\footnote{
This remark stems
from old unpublished work in collaboration with
Sudhir Chadha.}.
Very near the Fermi surface there should be at most two
relevant components, i.e. effectively n=2.

We will return to the issue about additional degrees of freedom for the
Weyl-spinor $\psi$ in section 3.6 where we
are able to offer ``An almost true theorem about the number of gauge
bosons and fermion components''. However, this theorem
is not fulfilled for
the left handed quarks (Weyl spinors) -  and
there is, in fact, some instability of the
left handed Weyl spinors (in the
restricted sense we have talked about here) when we take into account
the gauge degrees of freedom of the
Weyl spinor. But it is remarkable that all the {\em other} Weyl spinors,
i.e. the right handed quarks and the leptons, are stable (in our sense)
- provided you do not allow the basin of perturbations to allow
mixing of one irreducible representation with another one.   \\


\noindent
{\bf 3.} {\bf The basin of equations are chosen to be selected from
the class of differential equations}. \\
This restriction could be relaxed in many ways. One often
implement physics on a fine lattice (for instance with difference equations)
in the ultraviolet, which would lead to differential equations only as
``effective'' equations in the infrared. But that
would really only work if the ``lattice'' used
has enough structure to be approximately a manifold
from a long distance point of view. Even in the section 3
of H.B. Nielsen \& S.E. Rugh \cite{HolgerSvend}  where we attempt to make a
speculative model  which relaxes
many of the properties of the Weyl
equation we {\em only} relax the equation-operator
to be an analytic one in $\BFACE{x}$'s (or $\BFACE{q}$'s) and
$\BFACE{p}$'s or
equivalently differential operators
$\BFACE{p}= - i \partial / \partial \BFACE{x}$.
Since analytic functions are at least approximated by polynomials
it means that, approximately, we did only allow polynomials in
the differential operators and thus
essentially kept the requirement of
there being a {\em differential} operator in the
equation. We approximately kept a differential equation there.
Most important is presumably that we keep a kind of ``locality''
by keeping a differential equation. To attempt to relax this
assumption
might of course  be interesting, but very
likely this can only be done by  asserting (in the same go)
that in almost any structure you can invent an approximate topology
and thus pretend to see some locality.  \\

\noindent
{\bf 4.} {\bf The space-time dimensionality $d$ is kept fixed under the
perturbation $\delta W$ of the Weyl equation}    \\
Even if you allow a higher dimension of space(time)
there would be no $\sigma$ matrices to go with the
extra coordinates and momenta and one could show that
there would be no motion in the extra directions
of the modified Weyl particle.
There would, however, be
no obvious reason why a right handed top quark and a
right handed charm quark, say (i.e. fermions which belong to
different irreducible representations), should select the same directions
in which not to move. The story of ``no motion''
in more than 3 space directions is also
only true as long as the number of components of the Weyl spinor is
fixed to 2. However, we (at least) cite
suggestive arguments in H.B. Nielsen \& S.E. Rugh \cite{HolgerSvend})
for that genericly there will {\em effectively} only
be two components
(see also \cite{FroggattNielsen, Littlejohn}).


In particular - provided no mechanism prevents different
types of Weyl particles from selecting different ``directions
of no motion'' -  human observers, say,
composed of Weyl particles of these different
sorts, would be split up in subparts (moving in different
directions) each of which consists of
Weyl particles of a given type in the sense discussed above!
Also, there would be access to more than 3+1 dimensions
using several types of Weyl particles in this way.
\par
When a higher dimensional Weyl spinor with $d\geq 4$ can move ``fast'' in
more than three  spatial coordinate directions this is a property
that is strongly unstable (even
restricting
to a very small piece around a Fermi surface).   \\

We remark, that  every time we attempt to  relax one of the restrictions
1,2,3 and 4
mentioned above, we very fast end up with a class of models
which is infinitely much bigger than the restricted
basin of the defined perturbations.

\subsection{Preliminary discussion of some ``principial boundaries''
for which ``basin of perturbations'' we can formulate? }

We would like to state the ``stability'' of the Weyl equation as
impressive as possible. I.e. the Weyl equation would be stable
under ``all'' perturbations - conceivable as well as inconceivable.
This is, clearly, not possible to claim.
We cannot claim that the Weyl equation comes out from
{\em everything} (an elephant, say)
- even in some limit? Nevertheless \footnote{If
we take, say, the limit of extremely low temperature
the resulting ``deeply frozen'' elephant  might have left only a very tiny
fraction of its originally active degrees of freedom and it
is not so easy to exclude totally that the remaining vibrations
in the trunk and the still active spins
could not, somehow, be interpret cleverly as Weyl equation(s)
and/or the Standard Model ? }
this is the ``spirit in the random dynamics''!

Besides those restrictions of the basin just
mentioned above (item 1,2,3 and 4)
there may even be restrictions of a kind which
we humans may not be fully aware  -
and which may represent
principal boundaries
we may therefore neither be able to transcend nor fully explore,
even in
the distant future (and therefore we cannot easily display such
boundaries here).

If one should go so far as to even consider logic a part
of physics, and thus make a ``random physics'' model to be also
one which relaxes on the principles of our Aristotelian logic,
it would get very hard indeed to formulate the basin or class of
models of this type, let alone to prove something about that class.

The boundaries for what can be conceived or not are, surely,
very hard to locate.
As a simple illustration of a possible perturbation which you would hardly
imagine with ``old fashioned group theory'' is
the existence of ``quantum groups'' (e.g. $SU(2)_{q}$)
conceived of as a continuous deformation of a group.
If one should suggest a perturbation of the Weyl equation which
was ``analogous'' to such continuous deformation of groups
(``quantum groups'') we would have a very hard job to do! For
instance,  perturbation
of the equation into some ``relation'' among symbols which was not even an
equation.

However, it is - in our point of view - an interesting project to
{ identify},
and try to formulate in print,
such
fundamental boundaries and limitations for what can
- ever - be thought and accomplished.
For example, it is well known that
the german philosopher Immanuel Kant and the danish physicist Niels Bohr
were - seriously - contemplating
such issues.\footnote{Cf, e.g., I. Kant {\em ``Prolegomena to Any
Future Metaphysics''} (a shorter, more easily readable version
of {\em ``Critique of Pure Reason''}) \cite{Kant}.
Niels Bohr has discussed our ``Conditions for description of
Nature'' at very many occasions, cf. e.g. \cite{Bohr}. }

Niels Bohr, in particular, emphasized the important role played by our
``daily language'' which  presents a ``boundary'' in the sense that
it is hard to transcend - since we ultimately have to
communicate all our ideas and knowledge in ``daily language''.

We might add that any attempt to make ``great theories''
at a very fundamental level especially has to be
{\em accompanied}
with some {\em interpretation back} to the daily language.
Any such model is thus necessarily complicated by being
provided with such an unavoidable system of
interpretation
assumptions. Irrespective of how ``unified'' and
{ how} simple
some ambitious ``T.O.E.'' would be it cannot be
without some assumptions providing the translation from
its own formalism into that of daily language. This
fact sets a lower limit for its simplicity:
\begin{center}
We cannot
transcend that limit, but could we reach it?
\end{center}
You may say that the ``random dynamics'' project
precisely intends to reach it. The goal (in the strongest
form of ``random dynamics'') is to
argue that whatever the basic theory is - almost -
we can by appropriate interpretation,
i.e. by adding the appropriate ``translation assumptions'' into
daily language, make it become a good and well functioning
``T.O.E.'' But then it becomes also tempting to
really make use of this ``interpretation assumptions''
to the uttermost.

\subsection{So what could (should) ``random dynamics'' hope to
arrive at? }

Rather than claiming - as ideal ``random dynamics'' (a very strong form
of random dynamics
indeed
\cite{Scottish}) - that
{\em all} (or almost all \footnote{Both the concept ``all'' and ``almost''
are  not defined a priori.
(Presumably, they
could only be specified in a
way that would ultimately turn out to be rather arbitrary).
}) models at the fundamental
level lead to the well
known phenomenology (i.e. the Weyl equation, general relativity, the
standard model with its particular ``skew'' gauge group etc.)
we should seek
a {\em large class of models} that will lead to
the same infrared phenomenology.
The ``random dynamics'' project
intend to locate boundaries ``$ \partial {\cal B}$''
for this (huge?) class of models and
the project (however, at the level of ``Natural laws'')
resembles somewhat the project of finding ``universality classes'' in
phase transitions or other dynamical behaviors.

It is a natural contemplation, that if
these boundaries are not too narrow, it leaves (logically)
the possibility of having
{\em ``chaos in the fundamental laws''} - i.e. the ``fundamental structure''
may be selected randomly \footnote{And the
structure may thus very well differ - in a random way - in
various sub-parts
of our ``world machinery''. This idea is (in spirit) in some
contrast to the idea that there is only one unique
and universal pervasive ``superstring'', say, (a ``T.O.E.'')
which - at some fundamental scale -
is ubiquitously implemented ``everywhere''
- and implemented so ``precisely'' that it does
not differ, not even by the tiniest ``small error'',
in any sub-part of our ``world machinery''.
}
- if it is just selected from the class of models restricted by
these boundaries.

Thus, in some sense, there is a no need for finetuning
of the fundamental structure, except
that it has to lie within the boundaries.

Nevertheless, we note, that  these boundaries ``$\partial {\cal B}$ ''
(if these could be made precise enough to be stated formally, say, by
mathematical formulas)
in a certain sense themselves
have status as ``Natural Laws'' - and so on,
{\em ad infinitum}. Therefore, this way, one does not circumvent the concept
of some ``Natural laws''  to be implemented.

It would be interesting if the the ``random dynamics'' project
could lead to definitive conclusions as regards the
class of fundamental models
which would lead to the Standard Model in the infrared limit.
It may turn out to be of small (zero) measure
(in which case the fundamental structure have to be finetuned
enormously) or - which is more in the random dynamics spirit -
it may turn out that the models {\em not} leading to the standard model would
have small (zero) measure.

``Standard philosophy'', as materialized
in the last decades, for instance, in the search of
one unique ``superstring'' is inclined
towards the first viewpoint, while many arguments for the second viewpoint
have been collected in C.D. Froggatt \& H.B. Nielsen \cite{FroggattNielsen}.

Apparently, it is
easier to show ``instability'' than
``stability'', because the latter requires stability towards all conceivable
(and inconceivable, as well) perturbations, while the first
requires only the unambiguous demonstration of instability in some
directions of the conceivable perturbations.

If it turned out that the random dynamics project
failed in the sense that it could point towards some regularities
that has an instability - and thus these regularities
have to be finetuned - this
conclusion would in our opinion be a very interesting one, and point
towards some essential feature of the entire ``world machinery'' of which
we humans participate (so shortly).

In any case it is thus justified to {\em investigate}
if a finetuning or not is needed. Working the way that
one for every discovered law of Nature (which is based on empirical facts)
remembers to investigate its
stability could be conceived
of as using ``random dynamics ''
as a {\em scientific method}.

\subsection{Turning to the Weyl equation again}


In order to allow ourselves to call the Weyl equation stable, we
have - in addition to restricting the basin of perturbations (as discussed
above) - to diminish the ambition of stability by restricting
the features considered essential and therefore required to be
unchanged under the modifications (the perturbations).

Thus, when we claimed stability of the Weyl equation, under the considered
class of perturbations (\ref{Weylperturbation}), we only
achieved this because we claimed that it remained the Weyl equation
in spite of the fact that it actually was a completely new equation in the
sense that
the functional forms of the $A_{\mu}(x)$  and the $V_{\lambda}^{\mu}(x)$
had changed. It is only when we  maintain that these functional forms are not
essential for it being the Weyl equation that the Weyl equation
can be called stable, as we did.

This deemphasis  (needed for the stability) of the importance
of the functional forms of the $x$-dependent coefficients
$A_{\mu}$ (electromagnetic four potential) and $V_{\lambda}^{\mu}$ (vierbein),
is achieved, or strengthened at least, by declaring that they are
``dynamical''. By declaring  them ``dynamical'' we mean that
they depend on some bosonic degrees of freedom, say, so that  they
belong to initial conditions rather than to the laws of Nature.
They should be more like the position or the momentum of a particle
or the value of a field than like say the mass or the charge of
a particle or some coupling constant. Very luckily for such
a postulate of $A_{\mu}$ and
$V_{\lambda}^{\mu}$ being ``dynamical'' is the fact that
we happen to know phenomenologically already fields precisely having
the type of interaction with the Weyl particles which
$A_{\mu}$ and $V_{\lambda}^{\mu}$ have!
These fields are for $A_{\mu}$ the electric 4-potential and for
$V_{\lambda}^{\mu}$ the
vierbein, which is easily identified with a gravitational field.
In this sense we may say that we {\em need the electromagnetic
and the gravitational fields} in order to pretend the ``stability''
of the Weyl-equation! So instead of considering this
need for not considering the functional form essential
a weakness of the idea of claiming the Weyl equation exceptionally
stable in just 4 dimensions, it turns out to just fit
nicely provided we have electromagnetic and gravitational fields
(as dynamical fields). Rather than having a trouble
for our claim we got a postdiction of a couple of well established
forces in nature! \footnote{In the ``toy model''
which we put forward in section 3 of
H.B. Nielsen \& S.E. Rugh \cite{HolgerSvend}
we  construct (postulate)  a very general and
complicated quantum mechanical system (which we call
the ``bosonic'' part of the full model) interacting in
a random way with the ``fermionic world machineries''
which represented the Weyl particles. Since all the
effect of the ``bosonic part'' on the Weyl particles
is represented just by $A_{\mu}$ and $ V_{\lambda}^{\mu}$
it means that this ``bosonic part'' provides precisely
electromagnetic and gravitational fields. The model, however,
has a few ``small'' troubles: The interaction of the
vierbein and electromagnetic fields with themselves are
a priori non-local, we still have the dimensions
in which motion does not take place and, moreover, one
needs very strong
input assumptions about the state of the ``world machinery''
(the latter may, though, in principle be o.k.). }     \\


\noindent
{\bf Is the two dimensional Weyl equation really two
dimensional?} \\
Note, that if one imposes
the stability requirement (of the restricted
kind, we have discussed here), then all higher
dimensions than 4 have been completely ruled out, they are
not stable! Therefore, if ``our creator''
had chosen 6 dimensions say,
he had to ``finetune'' the Weyl equation, for example in order
to avoid terms  like some differential operator or some other
quantity multiplying a matrix not belonging to
the $\sigma$ matrices of the 6 dimensional Weyl equation!

If we  like to characterize{, however,} the dimension $4$
by the discussed stability
of the Weyl equation we meet the problem that the $d=2$ dimensional
Weyl equation is equally stable - or even more stable in a sense -
than the four dimensional one. For $d=2$ there are namely $n_{Weyl}=1$
component only and { hence} only $1^2=1$ linearly independent
matrices that can occur in the equation.

{}From the above formulas for the number of components it is easily seen that
the Weyl equation in $d=2$ dimensions has only $n=n_{Weyl}=1$ component
and thus we have in this case $d>n_{Weyl}^2$ and the Weyl equation must
even use the same ``$\sigma$-matrix'' more than once.
This means that it is
very stable, but if one have to use the same $\sigma$-matrix twice, one
can with good reason ask if this equation - the two dimensional Weyl
equation - is truly two-dimensional! We would say that it is really
not two dimensional, because the Weyl particle in two dimensional
Minkowskian theory, say, can only move in one direction and always with the
speed of light. It really only makes use of one light-cone variable, e.g.
$x^+=x^0+x^1$. Thus, it is rather to be considered only 1-dimensional!
If you accept this point of view a ``truly $d$-dimensional Weyl equation
which is stable in the sense described above must have equality between
the number of linearly independent $\sigma$ or $\gamma$ matrices $n^2$
and the number of dimensions $d$, i.e. $ n_{Weyl}^2=d $ for the Weyl case.

\subsection{An almost true theorem about the number of gauge bosons and
the number of different Weyl particles}


So far, we have considered only one single Weyl particle,
but we ought to consider (at least)
the states related by gauge transformations
as various components of
a single equation.
Taking into account that there can be several gauge components,
the number of components of the Weyl spinor increases\footnote{
This observation (obstacle) is due to a question by Prof.
Dragon at the conference}.
It is well known, that the Weyl spinor, indeed, has
such internal degrees of freedom (in Nature) -
although this has
been neglected in the previous discussion - as the reader
may have noticed.
For instance we know today that the Weyl spinor has color and
other gauge degrees of freedom. With such gauge fields interacting
with the Weyl particle the Weyl equation still takes formally the form
(\ref{WE})  but now the covariant derivative is {\em not} simply
\begin{equation}
 D_{\mu}=\frac{1}{i}\Omega(x,p)=\partial_{\mu}-ieA_{\mu}(x)
\end{equation}
(ignoring the vierbein by taking it to unit matrix), but
rather, say
\begin{equation}
 D_{\mu}=\frac{1}{i}\Omega(x,p)=\partial_{\mu}
-ie\frac{\lambda^a}{2}A_{\mu}^a(x)
\end{equation}
where the $\frac{\lambda^a}{2}$'s make up a basis in the
representation $r$ (for the Weyl particle in question) for
the representation of the Lie algebra of the Yang Mills gauge group. \\
I.e. there is one $\lambda^a$ matrix for each dimension of the
Lie algebra, and that again means one $\lambda^a$ matrix for
each ``color state'' of the gauge particle (each gauge particle
we could also say). In the Weyl-equation operator $\sigma^{\mu}D_{\mu}$
thus occurs the term
\begin{equation}\label{term}
\sigma^{\mu}  \frac{\lambda^{a}}{2} A_{\mu}^{a}.
\end{equation}
in the Yang-Mills case and
thus the Weyl field $\psi$ really must have
$n_{weyl} \cdot d_r$ components, where $d_r$ denotes the dimension
of the representation $r$ under which the Weyl particle transform
under gauge transformations. This representation dimension $d_r$
is
$$d_r=\left(  \begin{array}{c}
       \mbox{{\small number of different} } \\
       \mbox{{\small Weyl particles} }
         \end{array}
 \right) $$
while, as we remember,
$$n_{Weyl}=  \left(  \begin{array}{c}
       \mbox{{\small number of components} } \\
       \mbox{{\small of the  Weyl spinor} }
         \end{array}
\right).$$
The number of matrices that are needed to form a basis is
the square of the total number of components, meaning
$(n_{Weyl} \cdot d_r)^2$.
The condition for stability { of the Weyl equation (now coupled to ``color''
Yang-Mills degrees of freedom) - if we define it in an analogous manner
to the }  equation (\ref{observation}) above - is therefore given
by the following: \\

\noindent
\underline{``Almost true'' theorem}


\begin{eqnarray}  \label{almosttrueone}
 \left(  \begin{array}{c}
       \mbox{{\small number of different} } \\
       \mbox{{\small Weyl particles} }
         \end{array}
 \right)^{2}  & \times &
 \left(  \begin{array}{c}
       \mbox{{\small number of components} } \\
       \mbox{{\small of the  Weyl spinor} }
         \end{array}
\right)^{2} =  \nonumber \\
\left(  \begin{array}{c}
          \mbox{{\small number of } } \\
          \mbox{{\small gauge bosons} }
         \end{array}
\right) & \times &
\left(  \begin{array}{c}
         \mbox{{\small space-time} } \\
         \mbox{{\small dimension} }
         \end{array}  \right)
\end{eqnarray}
\par
\noindent
Here we have used that the number of different $A_{\mu}^a(x)$-components
is
$$ d_gd =   \left(  \begin{array}{c}
          \mbox{{\small number of } } \\
          \mbox{{\small gauge bosons} }
         \end{array}
\right)  \times
\left(  \begin{array}{c}
         \mbox{{\small space-time} } \\
         \mbox{{\small dimension} }
         \end{array}  \right), $$
This can be seen easily, since there is an $a$-value
for each
of the  $d_g$ dimensions of the gauge group.  \\


If we take into account that (at least empirically)
we have already the equation (\ref{observation}), our main observation
implies
\begin{equation} \label{a}
 \left(  \begin{array}{c}
       \mbox{{\small number of components} } \\
       \mbox{{\small of the  Weyl spinor} }
         \end{array}
\right)^{2} =  \nonumber \\
\left(  \begin{array}{c}
         \mbox{{\small space-time} } \\
         \mbox{{\small dimension} }
         \end{array}  \right).
\end{equation}\label{observationWheeler}
and combining this equation
with the more general stability condition (``the
almost true theorem'') we get :

\begin{equation}  \label{almosttruetwo}
 \left(  \begin{array}{c}
       \mbox{{\small number of different} } \\
       \mbox{{\small Weyl particles} }
         \end{array}
 \right)^{2} =
 \left(  \begin{array}{c}
       \mbox{{\small number of } } \\
       \mbox{{\small gauge bosons} }
         \end{array}
\right)
\end{equation}
Note, that this splitting of equation (\ref{almosttrueone}) into
(\ref{a}) and (\ref{almosttruetwo})
can be derived from the fact
that otherwise the matrices occurring in (\ref{term}) would
not be linearly independent.

So this is a prediction from the extension of the
observed stability of the Weyl equation.
Is it true empirically?

Had it not been for left handed quarks, this would have been an entirely
true prediction provided, though very strongly, that we (as
seen by Prof. Dragon) only considers
one single
irreducible representation at a time.
That is to say: For each irreducible representation
of Weyl particles - such as the right
handed strange quarks, say - we count the number
of gauge particles coupling non-trivially to this irreducible
representation and use {\em that} as ``the number of gauge particles''.
As the number of different Weyl particles we use
the number of different Weyl particles in the irreducible representation
in question.

{\em Then, indeed, the equation (\ref{almosttruetwo}) is
fulfilled for all the irreducible representations - except
for the left handed quarks!}
\par
For example, for the left handed positron
only one linear combination of gauge bosons couples
- that of the weak hypercharge - and there is
only one left handed positron in the irreducible representation,
so our equation is fulfilled with $1=1$.
Actually, in this case our previous discussion worked (in which
we neglected that there
is more than one component).

Another example is, say, one of the
left handed antiquarks: There are 3 different Weyl particles
in the irreducible representation, and to them couple
8 gluons and the weak hypercharge gauge boson (a certain
linear combination of $\gamma$ and $Z^0$). That is $ 8+1 = 3^2$.

Only for an irreducible representation of left handed quark we
are off: $12$ gauge bosons but $6$ Weyl particles. That does
not fit since $12 \neq 6^2$.


In fact it is so that for all other irreducible
representations than the left handed quarks one finds a true representation of
a $U(N)$ factor group of the standard model group $ SMG=S(U(2)\times U(3))$.
Here e.g. $N=1$ for the left handed positron, $N=2$ for the left handed
leptons , and $N=3$ for both types of left handed antiquark representations.
In such cases one finds
\begin{equation}
d_r=\left(  \begin{array}{c}
       \mbox{{\small number of different} } \\
       \mbox{{\small Weyl particles} }
         \end{array}
 \right)=N
\end{equation}
and
\begin{equation}
d_g=\left(  \begin{array}{c}
       \mbox{{\small number of } } \\
       \mbox{{\small gauge bosons } }
         \end{array}
 \right)=N^2
\end{equation}
so we would have the equality in just 4 dimensions $(3+1)$.
If we make the very important assumption of looking at only one
irreducible representation of the gauge group at a time we find that
indeed we have the stability in the standard model for all of them
except for the left handed quark representations.

So had it not been for
left handed quarks we could have
claimed that the stability of the Weyl equation
as used by Nature (i.e. the Standard Model)
not only supports the dimensionality
of space-time but also the type of gauge group representations found
in the standard model. Unfortunately for the achievements of this
principle of stability we have { an exception by the} left handed quarks,
or in other words: it would predict that there should not have been
left handed quarks, but otherwise it would have worked well {\em for each
irreducible representation separately}.

This latter lack of beauty means that the stability principle only is
o.k. provided there is some rule forbidding that different irreducible
representation of
fermions can get mixed in the attempts to modify the Weyl
equation(s).

The fact, that instability occurs when mixings are
allowed (here mixings between irreducible representations)
is also what we would see if we thought of the Dirac equation
instead of the Weyl equation. That would namely mean allowing
mixing of left and right handed components, and again no stability.



In conclusion, it appears that ``our creator'' has made
the Weyl equations
as stable as possible! This has been achieved both
by choosing the dimension
to be $4=3+1$ (and not, for instance, 8) and
by choosing the representations and the gauge group
so that in (at least) most cases each irreducible
representation taken separately is stable.

The exception of the
left handed quarks it is, in fact, rather unavoidable if there
have to be gauge anomaly cancellations
and we still want mass protected particles.
So, very likely, the choice of the Standard Model is - in some way
of measuring it - the most ``stable'' possibility.

So stability may even be a principle behind the representation choice, too!
Not only (as we already saw) behind the dimension and the existence
of gauge and gravitational fields.

It must be admitted, however, that there is the little problem
with the working of this principle now where we have included
several components and non-Abelian gauge fields:
We also get the vierbein and thereby the gravitational field
into a nontrivial representation of the gauge group.
That particular feature
is not so good from the empirical point of view!

\newpage

\section{Concluding remarks and dreams}

\setcounter{equation}{0}

In the present contribution we have made
an empirical observation (sec. 2.4) connected
to the Weyl equation,
suggested an interpretation in terms of ``form
stability'' of the Weyl equation (sec.3), and - in a search of
expanding the basin of allowed perturbations (modifications)
of the Weyl equation (under
which we would like to claim its stability) - we have
also challenged the concept of ``stability of Natural laws''.
The latter part (sec. 3.2-3.5) is a kind of preliminary attempt to locate
principal boundaries for the set of ideas which we call the ``random
dynamics'' project, described, e.g. in
C.D. Froggatt \& H.B. Nielsen \cite{FroggattNielsen}.

As concerns our empirical observation,  we
have seen that the dimension $d = 4$ obey the
equation (\ref{insertobservation}),
$$
d=\left\{\begin{array}{cccc}( 2^{d/2-1})^2 & = & 2^{d-2} &
\mbox{for $d$ even,}\\
(2^{(d-1)/2})^2 & = & 2^{d-1} & \mbox{for $d$ odd}.\end{array}
\right. \; \; \; ,
$$
which - via the step of the sigma-matrices
forming a basis (sec.3.1) - leads to the
suggestion that just in the phenomenologically true
dimension $d = 4$
is the Weyl equation especially stable.
This is so to say the message from
``our creator'' brought
via the dimension. In order to implement this stability it was at
first noticed that at low energy (and momentum) Taylor expansions had to
be used and - secondly - that we needed both gravitation and electromagnetism.
We may put it the way that in order to make
the idea of the Weyl equation being stable,
as suggested, one predict in retrospect these two forces of nature.
Really we could use the non-Abelian Yang Mills fields,
as found in the Standard Model,
instead of simple Abelian electrodynamics but there are a couple
of severe problems at this point:
\par
{\bf 1.} One must be satisfied
with looking only for stability under perturbations
in which fields in the same irreducible representation are allowed to mix, i.e.
we must require linearity and homogeneity in each irreducible representation
of Weyl fermions separately. (In the speculative model put up
in section 3 of H.B. Nielsen \& S.E. Rugh \cite{HolgerSvend}
one might hope
that mild continuity requirements will turn out to be enough for implementing
such a rule. But, a priori, it complicates matters to need such a separation
of the different irreducible representations).
\par
{\bf 2.} Even with the restriction to ``no mixing'' of different irreducible
representations the left handed quarks do not fit into the scheme.
The problem is that the entire standard model group has only dimension $12$
while the number of left handed quark Weyl fields is $6$ times the number
of components for a single Weyl two-componenet field. This means that
the number of allowed matrices is enhanced by a factor $6^2=36$ due to
the color and weak isospin degrees of freedom. The number of Yang-Mills
field components is, however, correspondingly only enhanced by $12$,
and that is less than $36$. In order that the ``form stability'' of
the Weyl equation shall work for the left handed quarks it
would be appropriate with a $U(6)$-gauge theory rather than, just, the
Standard Model.
Such a model would, however, have gauge anomalies and
not have a consistent gauge symmetry.
\par
We presume, that a  strongly related expression for the special ``form
stability'' of the Weyl equation
is the fact that just in the dimensions
$d = 3$ and $d=4$  is the number of helicity states
of a Weyl {\em particle} (antiparticle not included)
precisely one, so that there is {\em no degeneracy}.
\par
In both  ways of looking at our
result there is a need for a separate discussion of the
two dimensional ($d=2$) case\footnote{The ``no degeneracy''
principle leaves also $d=3$ as a competitor
to the experimental dimension $d=4$, but if
we assume that we only have access to low
energies and thus only shall see mass-protected
particles we could on the that ground exclude the
odd space-time dimensions anyway.}
since this is, at first,
even  more stable than the experimentally observed case
$(d=4)$. However, one may claim that this case
is not truly two-dimensional since it does not (truly) make use
of  both dimensions (cf. section 3.5).
\par
Taking (in spite of the ``small''troubles) the special stability
of the $4$-dimensional Weyl equation
as a trace in the search for fundamental principles
in physics we  have then looked into
how much we may {\em expand} the
basin of allowed perturbations
(modifications) and still claim the Weyl equation to
remain ``form stable''.
In a sense, such a search is, precisely, what is done
in H.B. Nielsen \& S.E. Rugh \cite{HolgerSvend}
in which we have put up a very general
``toy-model'' for a pregeometry (a ``world machinery'').
The latter type of model may in fact be thought of
as representing a very general type of linear homogeneous
{\em differential} equation conceivable as describing a rather
general quantum mechanical ``machinery'' having partly a classical
analogue and obeying the smoothness conditions associated with
the smoothness properties of this classical analogue.
It must, however, be admitted that
e.g. Fermi statistics and the identical particle property is rather put
in than derived, so far.
\par
Even if there are several troubles
(as we have described, sec. 3.2-3.6)
in interpreting the Weyl equation
as totally stable against all imaginable perturbations, we
would still claim
that the choice of just 4 dimensions, indeed, points to some sort of stability
as a guiding principle.

``Our creator'' has also
chosen the representations in the standard model so that
the Weyl equations become as stable as possible ! (at least very stable
compared to what could have happened). When He namely use
- as is the case except for the left handed quarks -
defining representations  of an $SU(N)$
group having also a nontrivial Abelian charge
there are $N^2$ gauge fields each with d = 4 Lorentz
components and also 4 times $N^2$ linearly independent
matrices. Only for the left handed quarks for which
we get 6 times 2 components there are 144 linearly independent
matrices but only 12 times 4 gauge potentials to adjust
(i.e. $48 \neq 144$).

It turns out that there is a conflict
between making the Weyl equation stable and
keeping mass-protection
(i.e. unless we satisfy the no anomaly condition trivially).
For instance, if one
simply postulated some gauge fields in a group SU(6) extending the
group $ SU(2) \times SU(3)\subseteq  SU(6)$ providing all the (special) unitary
transformations of an irreducible representation corresponding to
 a set of left handed quarks this field would have anomalies
in the gauge symmetry. I.e. the gauge symmetry could not  be uphold at
the quantum level unless further particles
were added to the model to compensate the anomaly.
But then, of course, the added field might be
in danger of lacking stability again.

Hence, a claim that there should be enough gauge fields to realize the
``form stability'' proposed seems  not
to be fully realized in Nature
- although Nature comes close - but it is also not realizable with
mass protected fermions: One simply cannot cancel the gauge anomalies and
at the same time have all the fermions in the defining representations
unless we choose a trivial cancellation which will spoil the mass-protection.
With so many defining representations as are used in
the Standard Model
one might then think that the Standard Model does its best possible to
be maximally ``stable''!\footnote{In fact, what favours the stability
is that the representation is so ``small'' as possible compared to the ``size''
of the gauge group or rather the part that is not represented trivially.
So maximal stability of the Weyl equation would roughly
correspond to smallest possible representations of the fermions.
Now it happens that S.Chadha and one of us pointed out that
the Standard Model representations are characterized as the ``smallest''
possible ones with mass-protection and no gauge-anomalies (we thank J. Sidenius
for finding that a single alternative solution is excluded by the requirement
of no mixed anomaly). So it is from that work really suggested that
the Standard Model is as stable as it would have been possible to
construct it without loosing  mass-protection and/or
the anomaly cancellations.}

\subsection{Could our observation be an accident? }

 Which ``signals'' in phenomenology carry much information and which
``signals'' carry only small amounts of information ?  \\
The purpose of reading out ``signals'' from phenomenology is of course
that this is the way we may really learn from Nature rather than
only from our own speculations. It is, however, only
to be trusted
when the number explained is sufficiently complicated that it is
not too easy to find an explanation for it.
There must be a fair amount
information
(which have previously not been understood)
that finds an explanation, otherwise
the explanation will not be convincing.

Moreover, there can
{\em only be one} explanation of a given phenomenon - such as
the dimensionality of space-time considered here. If there were
two fundamental (independent) explanations of a given
interesting number
(such as $4 = 3+1$ dimensions) then it would be a strange mathematical
accident that these two explanations would give the same result.

In that sense fundamentally different
attempts of arriving at the number 4 (for instance, the
alternative attempts \cite{Gibbons} - \cite{Antoniadis}
mentioned in the list of references) cannot be true -
provided, for instance, that our observation is
not an ``accident''.\footnote{That our observation is not an
accident would be strongly supported if we could obtain from
the Weyl equation
- in a related way - not only the dimension 4 but
also its splitting into 3+1.
The Weyl equation supplemented
with the - rather mild and reasonable  - assumption
that by appropriate multiplication with a matrix the
Weyl-equation operator $ \sigma^{\mu}D_{\mu}$ becomes {\em Hermitean}
can actually
only be realized provided the splitting is into an odd number of times
(and an odd number of space-dimensions; we think of the even $d$ case).
A relatively easy series of algebraic manipulations shows this
fact. For the troubles (as far as the Dirac or
Weyl equations are concerned)
caused in ``euklidianization'' due to the
difference between ``4'' and ``3+1'',
see e.g. S. Coleman \cite{Coleman}.
}

Note, that the dimensionality of space-time is known with
several digits precision,
$$ d = 4 = 4.00 .... $$
But almost all these digits are ``0''s, and therefore they do not
carry so much information - modulo that most underlying theories,
by construction, have an integer number of dimensions.

It would be easier to get reliable
information from the electromagnetic finestructure constant, say,
\footnote{Cf. ,e,g., {\em ``Review of Particle Properties''},
Phys. Lett. {\bf B 204} (April 1988).   }
$$ \alpha = \frac{e^2}{\hbar c} =
1/(137.0359895 \pm  0.0000061) $$
which is
such an {\em ``irregular''} number
(measured to a precision of 8 digits or so)
that it probably carries a huge amount of information to support or
destroy a theory which aims at predicting this
constant (we remark that no theory exist today which claims to arrive at
$\alpha$ with 8 digits precision\footnote{Actually it happens that
one of us would claim to have quite an impressive ``explanation'' of
all the three finestructure constants in the Standard Model by
relating them to the (or a) {\em ``multi-critical point''} (where
several phases of a - lattice - gauge theory meet)
But these fits have rather only one significant digit,
although that seems to us already suggesting that there is
some truth behind. Cf. D.L. Bennett \& H.B. Nielsen, in preparation.}).

Compare, also, e.g. with
the Standard Model group $S(U_2 \times U_3)$.
If we want to predict this group, relative to other possible groups
with up to 12 generators (gauge bosons), we have of the order of 240
groups \cite{skewness}. We conclude that to predict
the true gauge group is
comparable to predicting  a number with two significant digits.  \\

If we, by accident, find several fundamentally different
explanations for the observed dimensionality $d = 4$ of our space-time
then it would presumably
indicate that the number ``4'' is too small and too simple (there is not
encoded information enough in this number) for reading out
interesting structural information from it.
In the light of that several explanations - which seems reasonable and
fundamental, e.g. the stability of (classical or quantum)
systems governed by the Coulomb or Newtons law
- coexist, we fear that the number ``4''
is too simple to be readable in this sense.

\newpage

\section*{Acknowledgements}

We both have pleasure in thanking for travel-support the EEC-grant CS1-D430-C
which has been of significance for developing the ideas of the present article
for instance in Rome and Heraklion and in Berlin, where the conference was
held.

S.E.R. would also like to thank Prof. John Negele and Prof. Ken Johnson
for the warm hospitality extended to him at the Center for Theoretical
Physics, where parts of this work was written up in
some final stages.

Support from the Danish Natural Science Research Council
(Grant No. 11-8705-1) is gratefully acknowledged.

We thank Hans Frisk for discussions and important input at a certain
stage of the development of this work, in particular, discussion about
the ``mode conversion surfaces'', a central concept for the
toy-model of a ``world machinery'' described in a
previous work \cite{HolgerSvend} on these issues.


\addcontentsline{XXX}{section}{References}

\begin{thebibliography}{X}

\bibitem{BarrowTipler}
J. D. Barrow \& F. J. Tipler,
{\em ``The Anthropic Cosmological Principle''},
Clarendon Press, Oxford (1986), especially chapter 4.8.  \\
J.D. Barrow {\em ``Dimensionality''} Phil.Trans.Roy.Soc.Lond.A.
{\bf 310}, 337 (1983).

\bibitem{Ehrenfest}
P. Ehrenfest {\em ``Physical Laws and Dimensions of Space''}
Proc. K. Acad. Amsterdam. {\bf 20}, 200-209 (1917).

\bibitem{Whitrow}
G.J. Whitrow {\em ``Why Physical Space Has Three Dimensions''},
Br. J. Philos. Sci. {\bf 6}, 13-31 (1955).


\bibitem{Scottish}
H.B. Nielsen, {\em ``Dual Strings'' - Section 6. Catastrophe Theory Programme},
{\em in:} I.M. Barbour \& A.T. Davies (eds.),
{\em ``Fundamentals of Quark Models''}, Scottish Univ. Summer
School in Phys. (1976) pp. 528-543.   \\
H.B. Nielsen {\em ``Field theories without fundamental gauge symmetries''},
Phil.Trans.Roy.Soc.Lond.A. {\bf 320}, 261 (1983).


\bibitem{FroggattNielsen}
C.D. Froggatt \& H.B. Nielsen, {\em ``Origin of Symmetries''},
World Scientific (1991) and references (and reprints) therein, especially
section 7.2.2.B. For the argument by S. Chadha and H.B.N. for
there being effectively only two components at the Fermi surface
see page 146, in this section.


\bibitem{FuNielsen}
Y.K. Fu \& H.B. Nielsen, Nucl.Phys. {\bf B 236} (1984) 167-180;  \\
Y.K.Fu \& H.B. Nielsen, {\em ``On random dynamics and layer phase -
A new way of dimensional reduction giving hope of understanding an
origin of topology of space time''}, contr. to the
Proc. of the XVII Int. Symp., Ahrenshoop, DDR, Oct. 1983.

\bibitem{skewness}
H.B. Nielsen \& N. Brene, Phys. Lett. {\bf B 223} (1989), 399; \\
H.B. Nielsen \& N. Brene {\em ``Skewness of the Standard Model -
Possible Implications''}, in P. Nicoletopoulos and J. Orloff
``The Gardener of Eden'' (In Honour of Robert Brout for his
60 years birthday), Physicalia Magazine Vol. {\bf 12} (1990).   \\
H.B. Nielsen \& N. Brene {\em ``Splitting of the weak hypercharge
quantum''}, Nucl. Phys. {\bf B 359} (1991), 406-422.

\bibitem{HolgerSvend}
H.B. Nielsen \& S.E. Rugh {\em ``Weyl Particles, Weak Interactions and
Origin of Geometry''}, Nucl.Phys.{\bf B} (Proc.Suppl.)
{\bf 29 B,C} (1992), 200-246.

\bibitem{SvendetalNOGO}
S.E. Rugh et al. {\em ``Towards a (Simple) No Go Theorem
for a ``Theory of Everything''}, in preparation.


\bibitem{Littlejohn}
R.G. Littlejohn \& W.G. Flynn, Chaos {\bf 2} (1992) 148. \\
L. Friedland, Phys. Fluids {\bf 28} (1986) 3260.



\bibitem{Gibbons}
G.W.Gibbons {\em ``The Dimensionality of Spacetime''} in H.J.de Vega and
N. S\'{a}nchez (eds.) {\em ``String Theory, Quantum Cosmology and
Quantum Gravity. Integrable and Conformal Invariant Theories''}
(World Scientific, 1987).

\bibitem{Maeda}
K. Maeda {\em ``The Einstein Gravity as an Attractor in Higher-Dimensional
Theories''}, p.426-433 in H. Sato et al. (eds.) {\em ``Gravitational
Collapse and Relativity''} (World Scientific, 1986).

\bibitem{Demianski}
A. Chodos \& S. Detweiler, Phys. Rev. {\bf D 21}, 2167 (1980) (purely
dynamical reduction). M. Szydlowski,
J. Szczesny \& M. Biesiada, Class.Quant.Grav. {\bf 4}, 1731 (1987);
M. Demianski, M. Heller \& M. Szydlowski, Phys.Rev.{\bf D 36}, 2945 (1987).

\bibitem{Antonsen}
F. Antonsen {\em ``Pregeometry''}, Cand.Scient.Thesis, The Niels Bohr
Institute (1992); Second revised edition with small changes
(September 1992), distributed.  \\
F. Antonsen {\em ``Comment on Quantum Metric Spaces''},
Niels Bohr Institute preprint, NBI-HE-92-37.

\bibitem{Alvarez}
E. Alvarez, J. Cispedes \& E. Verdaguer {\em ``Quantum metric spaces as a
model for pregeometry''}, Phys.Rev.{\bf D 45} (1992), 2033.


\bibitem{HochbergWheeler}
D. Hochberg \& J.T. Wheeler {\em ``Spacetime dimension from a variational
principle''}, Phys.Rev.{\bf D 43}, 2617 (1991).

\bibitem{Greensite}
J. Greensite {\em ``Dynamical Origin of the Lorentzian Signature of
Spacetime''}, Niels Bohr Institute preprint, NBI-HE-92-59.

\bibitem{Penrose}
R. Penrose, {\em ``Twistor theory: its aims and achievements''},
{\em in:} C.J. Isham, R. Penrose and D.W. Sciama (eds.),
{\em ``Quantum Theory''}, an Oxford Symp., Oxford Univ.
Press (Oxford, 1975).


\bibitem{strings}
M.B. Green, J.H. Schwarz \& E. Witten {\em ``Superstring Theory''}
(Cambridge Monographs on Mathematical Physics, Cambridge
University Press, 1987). In two volumes.

\bibitem{Vafa}
R. Brandenberger \& C. Vafa {\em ``Superstrings in the Early Universe''},
Nucl. Phys. {\bf B 316} (1989), 391. \\
A.A. Tseytlin \& C. Vafa, {\em ``Elements of String Cosmology''},
Nucl. Phys. {\bf B 372} (1992), 443.

\bibitem{Antoniadis}
I. Antoniadis {\em ``A possible new dimension at a few TeV''},
Phys.Lett.{\bf B 246}, 377 (1992) and references therein. \\
H.B. Nielsen and P.H. Frampton, in preparation.


\bibitem{Weinberg}
S. Weinberg {\em ``Conference Summary''}. Talk presented at the XXVI
International Conference on High Energy Physics, Texas, August 12 (1992),
University of Texas (Austin) preprint, UTTG-25-92.


\bibitem{Coleman}
S. Coleman, {\em ``Aspects of Symmetry''}, Selected Erice Lectures,
Cambridge Univ. Press (1985).


\bibitem{Lehtoetal}
M. Lehto, H.B. Nielsen \&  M. Ninomiya,
{\em ``Time translational symmetry''},
Phys. Lett. {\bf B 219} (1989) 87-91
and references therein.  \\
M. Lehto, H.B. Nielsen \& M. Ninomiya,
{\em ``Semilocality of one-dimensional simplicial quantum gravity''}
Nucl.Phys. {\bf B 289}(1987) 684-700
and references therein.

\bibitem{AtleeJackson}
E. Atlee Jackson {\em ``Perspectives of nonlinear dynamics''},
(Cambridge University Press, 1989), Volume 1,
especially pages 41, 103.  \\
V. Szebehely {\em ``Review of Concepts of Stability''},
Celestial Mech. {\bf 34}, 49-64 (1984).

\bibitem{ColeyTavakol}
A.A. Coley \& R.K. Tavakol {\em ``Fragility in Cosmology''},
Gen.Rel.Grav. {\bf 24}, 835 (1992) and references therein.

\bibitem{Kant}
I. Kant {\em ``Prolegomena to Any Future Metaphysics''} (1783)
(The Paul Carus Translation extensively revised by J.W. Ellington is
available from Hackett Publ. Co, 1977).
This little book affords the reader a compact (and more accessible)
overall view   of {\em ``Critique of Pure Reason''} (1781).


\bibitem{Bohr}
N. Bohr {\em ``Atomic Theory and the Description of Nature''}
(Cambridge, 1934), {\em ``Atomic Physics and Human Knowledge''}
(New York, N.Y., 1958), {\em ``Essays 1958-1962 on Atomic Physics and
Human Knowledge''} (New York, N.Y., 1963).


\end{thebibliography}
\end{document}